\documentclass[letterpaper,twocolumn,10pt]{article}
\usepackage{usenix2019_v3}

\usepackage{tikz}
\usepackage{amsmath}
\usepackage{filecontents}
\usepackage[linesnumbered,ruled]{algorithm2e}
\usepackage{url,textcomp}
\usepackage{float}

\def\Section {\S}

\newcommand{\squishlist}{
 \begin{list}{$\bullet$}
  { \setlength{\itemsep}{0pt}
     \setlength{\parsep}{3pt}
     \setlength{\topsep}{3pt}
     \setlength{\partopsep}{0pt}
     \setlength{\leftmargin}{1.5em}
     \setlength{\labelwidth}{1em}
     \setlength{\labelsep}{0.5em} } }

\newcommand{\squishlisttwo}{
 \begin{list}{$\bullet$}
  { \setlength{\itemsep}{0pt}
     \setlength{\parsep}{0pt}
    \setlength{\topsep}{0pt}
    \setlength{\partopsep}{0pt}
    \setlength{\leftmargin}{2em}
    \setlength{\labelwidth}{1.5em}
    \setlength{\labelsep}{0.5em} } }

\newcommand{\squishend}{
  \end{list}}

\begin{document}

\date{}

\title{\Large \bf No Reservations: A First Look at Amazon's Reserved Instance Marketplace}
\author{\rm Pradeep Ambati, David Irwin, and Prashant Shenoy\\University of Massachusetts Amherst}

\maketitle

\begin{abstract}
Cloud users can significantly reduce their cost (by up to 60\%) by reserving virtual machines (VMs) for long periods (1 or 3 years) rather than acquiring them on demand. Unfortunately, reserving VMs exposes users to \emph{demand risk} that can increase cost if their expected future demand does not materialize. Since accurately forecasting demand over long periods is challenging, users often limit their use of reserved VMs.  To mitigate demand risk, Amazon operates a Reserved Instance Marketplace (RIM) where users may publicly list the remaining time on their VM reservations for sale at a price they set.  The RIM enables users to limit demand risk by either selling VM reservations if their demand changes, or purchasing variable- and shorter-term VM reservations that better match their demand forecast horizon.   Clearly, the RIM's potential to mitigate demand risk is a function of its price characteristics.  However, to the best of our knowledge, historical RIM prices have neither been made publicly available nor analyzed. To address the problem, we have been monitoring and archiving RIM prices for $1.75$ years across all 69 availability zones and 22 regions in Amazon's Elastic Compute Cloud (EC2).  This paper provides a first look at this data and its implications for cost-effectively provisioning cloud infrastructure.

\end{abstract}
    
\section{Introduction}
\label{sec:introduction}
Cloud platforms, such as Amazon's Elastic Compute Cloud (EC2), enable users to reserve virtual machines (VM) for either 1 or 3 years for a substantial discount $d$ of $40$\% or $60$\%, respectively, off the on-demand price per unit time~\cite{reserved-instances}.  Thus, as long as the percentage of time cloud applications use reserved VMs is high, i.e., much greater than ($1-d$), they are cheaper than using on-demand VMs over the same time period.   In addition, unlike on-demand VMs, which may be unavailable during peak demand periods, reserved VMs provide a capacity reservation that assures users they will always be available if required over the reservation's term.


Of course, if reserved VMs' utilization is low, i.e., much less than ($1-d$), then they are substantially more expensive than dynamically acquiring and releasing on-demand VMs only when necessary. Thus, the optimal number of reserved VMs to provision to minimize cloud costs is a function of their expected utilization over the reservation's term. Unfortunately, accurately forecasting computing demand over long multi-year periods is challenging, as both technology and the economy change on much shorter time scales.  For example, unforeseen events, such as a global pandemic, can substantially change computing demand.  As a result, reserving VMs exposes users to substantial \emph{demand risk}, or the risk of losses due to the gap between forecasted and actual demand.  


To mitigate demand risk, Amazon operates a Reserved Instance Marketplace (RIM) where users may publicly list the remaining time on their VM reservations for sale at a price they set~\cite{rim}. The RIM enables users to limit demand risk by either selling VM reservations if their forecasted demand changes, or by purchasing variable- and shorter-term VM reservations that may better match the horizon over which they can accurately forecast their demand. Clearly, the RIM's potential to mitigate demand risk is a function of its price characteristics.  However, to the best of our knowledge, prior work has not analyzed RIM price data to understand its effect on cost-efficient cloud provisioning.  While some prior work has examined using the RIM to optimize cloud costs, it is analytical and does not consider the effect of real RIM price data~\cite{rim1,rim2,rim3,rim4}. Instead, this work uses the RIM as inspiration for defining and analyzing new problems in online algorithms. 

The RIM differs in important respects from EC2's spot market, which offers spare computing capacity for a variable price and has been well-studied~\cite{deconstruct,bhuvan-spot2}.  In particular, the RIM is a competitive market that brings together multiple buyers and sellers, where sellers set their own price. In contrast, EC2's spot market includes only a single seller (Amazon) that sets the price based on a hidden pricing algorithm that frequently changes and is not market driven.  As a result, the RIM likely represents the only large-scale competitive market for computing resources with dynamically changing prices set based on supply and demand.   In addition, the spot price for different VM types in each availability zone (AZ) are uniform and change in real time, while users list VM reservations on the RIM for a wide range of prices. These listings also have a wide range of characteristics that affect their price, such as the number of VMs in the reservation, their payment type, and their remaining term.  Thus, while the spot market is akin to the stock market---with uniform pricing of many identical assets---the RIM is more akin to the housing market with many listings with different characteristics that affect price. 


While the RIM is a much richer and more complex marketplace than the spot market, it is also much more opaque.  In particular, EC2 \emph{does not} automatically archive RIM price data and make it accessible for users, as it does for spot prices, which are always available for the past 3 months.  While many third parties maintain spot price data archives going back many years, we could find no similar online archives of RIM price data, even though the RIM has been operating since 2012~\cite{rim-start}.  Thus, there appears to be no historical RIM price data available to enable users to make informed decisions as to how to use the RIM for long-term cloud provisioning, or how to price their VM reservations in the market.  To address the problem, we have been monitoring and archiving RIM prices for $1.75$ years across all 69 AZs and 22 regions in EC2.

This paper provides a first look at RIM price data and its implications for cost-effective cloud provisioning.  \Section\ref{sec:background} details the different reserved VM pricing options and the RIM's basic rules, both of which embed some complexity that affects prices.  \Section\ref{sec:design} then analyzes RIM prices from 2018/9 to 2020/5 along multiple dimensions, including market volume, listing duration, price, and time on the market.  We use our analysis to make multiple insights, some of which we highlight below.

\squishlist
\item The RIM is large with tens of thousands of listings and average monthly sales in the millions of dollars. 
\item There are substantial differences in RIM market volume between regions, across time, and among VM types. 
\item Market volume is still small enough that, even in the largest regions, a single user can significantly affect it.
\item Regions with larger market volume tend to have longer average time-on-the-market indicating lower demand.
\item There are opportunities for savings due to price inversions where some prices are much less than Amazon list prices. 
\item There is evidence of mispricing on listings where some prices are much higher than Amazon list prices. 
\item Most reservations are short-term (1-6 months) and most VM types are compute-optimized or general-purpose. 
\item The RIM enables users to reserve VMs for as little as 1 month at an effective price similar to that of the spot price. 
\squishend

We intend our analysis as a first step in leveraging the RIM to cost-effectively provision cloud infrastructure, and many of our insights deserve additional attention to understand how best to exploit them for cost savings.  To encourage further work, we have made our dataset publicly available at the UMass Trace repository (\url{http://traces.cs.umass.edu}).

 \section{Reserved Instance Marketplace Rules}
 \vspace{-0.1cm}
 \label{sec:background}
 We provide some background on Reserved Instances (RIs), the RIM's market rules, and the data we have collected.


\noindent {\bf Reserved Instances}.  While EC2 offers different types of RIs, we focus on standard RIs since they offer the highest discount and \emph{are the only type users can sell in the RIM}. EC2 offers standard RIs for 1 and 3 year terms at a 40\% and 60\% discount, respectively, off the on-demand price per unit time over the term's length.  An RI's utilization determines its \emph{effective discount}.  The discounts above only apply if the RI is utilized 100\% of the time.  If an RI is utilized $u$$<$$100$\% of the time and its full discount is $d$, then its effective discount compared to using an on-demand VM when necessary is $u \times (1-d)$. 

Users may also select from three different RI payment options: all upfront, partial upfront, and no upfront.   The all upfront option requires users to pay the entire cost of the RI upfront, and offers the largest discount, i.e., the 40\% or 60\% discounts above.  Partial upfront requires users to pay $\sim$50\% of the all upfront price plus a monthly cost, and offers a slightly lower discount, at 1-2\% less than all upfront, over the reservation's term. Similarly, the no upfront option requires no upfront payment, but at a higher monthly cost, and offers an even lower discount, at 3-4\% less than all upfront cost. 

\noindent {\bf RIM Market Rules}.  The RIM includes many rules and restrictions that affect its prices and operation.  Importantly, while users can list all upfront, partial upfront, or no upfront RIs, they can only sell the upfront portion, with the purchaser responsible for any remaining monthly payments.   Thus, the vast majority of no upfront RIs are listed for \$0, as the buyer only purchases the capacity reservation and the responsibility for making the monthly payment.  There is no explicit provision against re-selling RIs purchased in the market.  However, Amazon takes a 12\% service fee from the seller on each transaction, and mandates a 30-day holding period before listing any RI in the RIM, both of which limit speculation. 



Sellers set their own price when listing RIs on the market, although Amazon suggests a selling price based on the reservation's remaining term and the market's current price for a similar VM type and term.  For uniformity, Amazon truncates all remaining terms to the month. Buyers then request a specified number of VMs of a certain type and term (in months).  Amazon groups RIs based on their remaining term and upfront price, and satisfies buyer requests by selling RIs in order of the lowest upfront price with the specified term.  Note that Amazon may fulfill requests from multiple different sellers, and may split sellers' listings by only selling some of its RIs.  Each listing has a unique public identifier that lists its attributes, e.g., VM and reservation type, quantity, term, upfront price, etc., which enables us to track when listings change due to a partial sale.  As we show in \Section\ref{sec:design}, listings are often on the market for multiple months, although users may still use the RIs after they are listed but not yet sold. 




\noindent {\bf Data Collection}. We wrote simple python scripts using EC2's Boto3 API to collect RIM data starting in September 2018.  Specifically, we call {\tt describe-reserved-instances-offerings} every 30 minutes for every VM type in every AZ of every region.   We record each field in the API response, such as the RI offering identifier, duration, fixed price, usage price, VM type, AZ identifier, and quantity. Note that we cannot track actual sales of RIs versus listing cancellations, but can only observe when listings come on and go off the market.

\vspace{-0.1cm}
\section{Market Data Analysis}
\label{sec:design}
\begin{figure}[t]
\centering
\includegraphics[width = 0.5\textwidth]{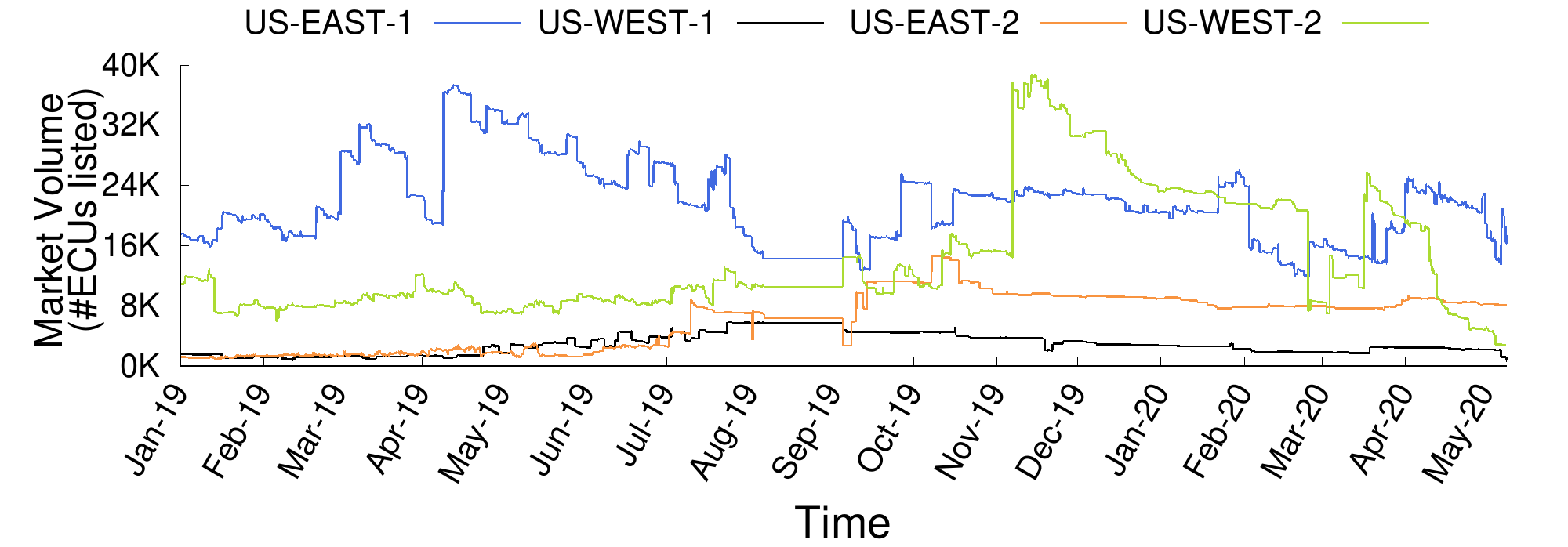}
\vspace{-0.7cm}
\caption{\emph{RIM volume in ECUs for different U.S. regions.} }
\vspace{-0.3cm}
\label{fig:aggregate}
\end{figure}

\begin{figure}[t]
\centering
\includegraphics[width = 0.5\textwidth]{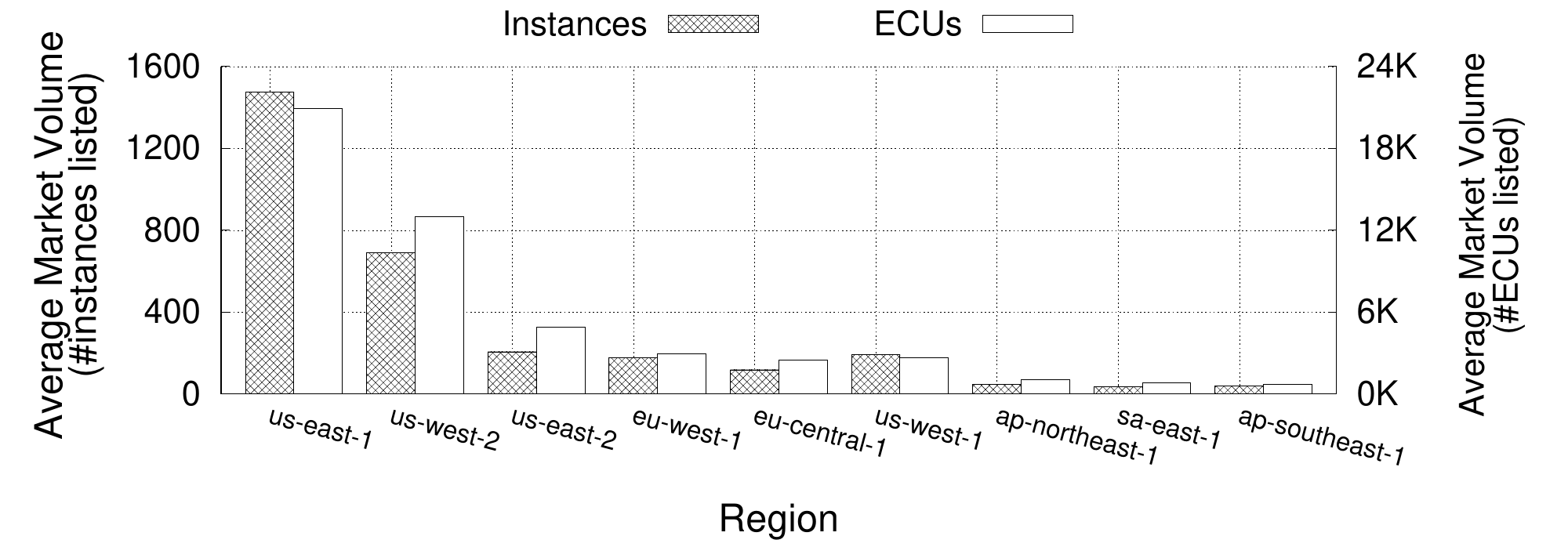}
\vspace{-0.9cm}
\caption{\emph{Average volume across different regions.}}
\vspace{-0.5cm}
\label{fig:compare-regions}
\end{figure}

We analyze RIM price data over $1.75$ years from September 2018 to May 2020 to gain insights into using it to cost-effectively provision cloud resources.  Since our data reflects Amazon's cloud, which is large and complex---with dozens of VM types, regions, AZs, and reservation terms---we isolate and focus on specific aspects of the market's data below. 

\noindent {\bf Aggregate Market Volume}. Figure~\ref{fig:aggregate} plots the market volume over time in terms of the number of ECUs listed.  Amazon defines an ECU as a relative measure of a VM's integer processing capacity  This graph aggregates all listings and VM types on the market, and gives a sense of the relative size and scope of the market in a few different regions.  We focus on the U.S. regions, since they are the largest and most mature. 


\begin{figure}[t]
\centering
\includegraphics[width = 0.5\textwidth]{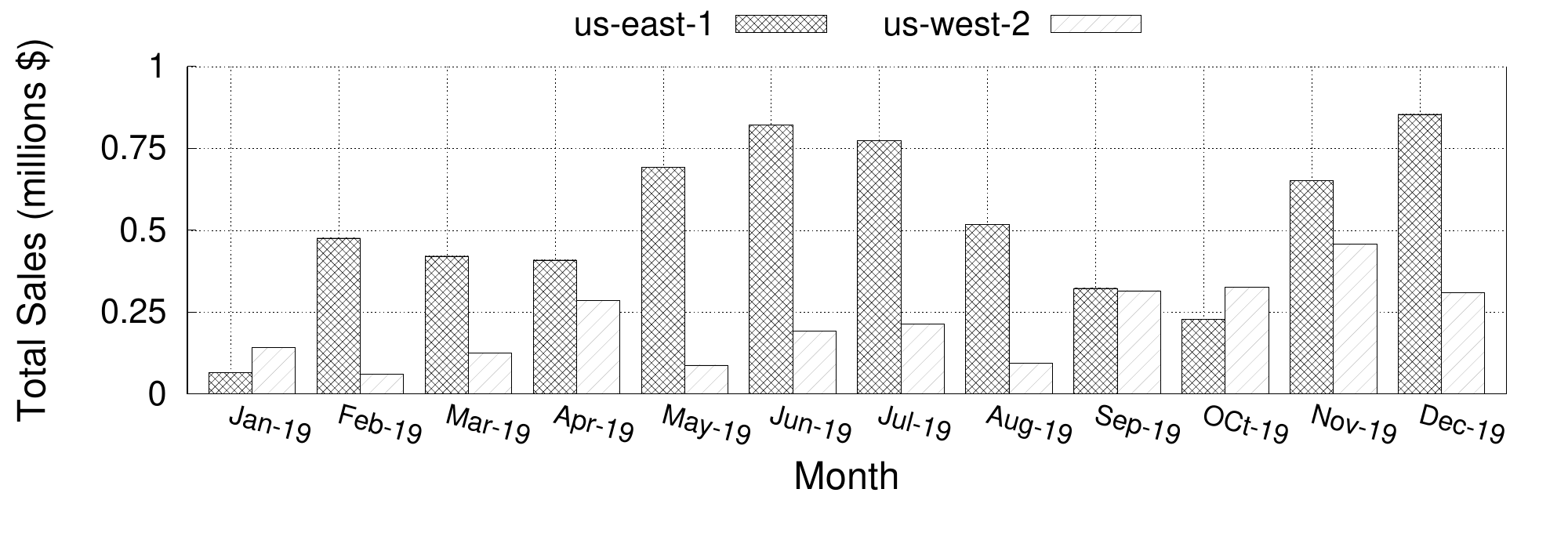}
\vspace{-0.8cm}
\caption{\emph{Monthly RIM sales for us-east-1 and us-west-2.}}
\vspace{-0.4cm}
\label{fig:monthly-sales}
\end{figure}

\begin{figure}[t]
\centering
\includegraphics[width = 0.5\textwidth]{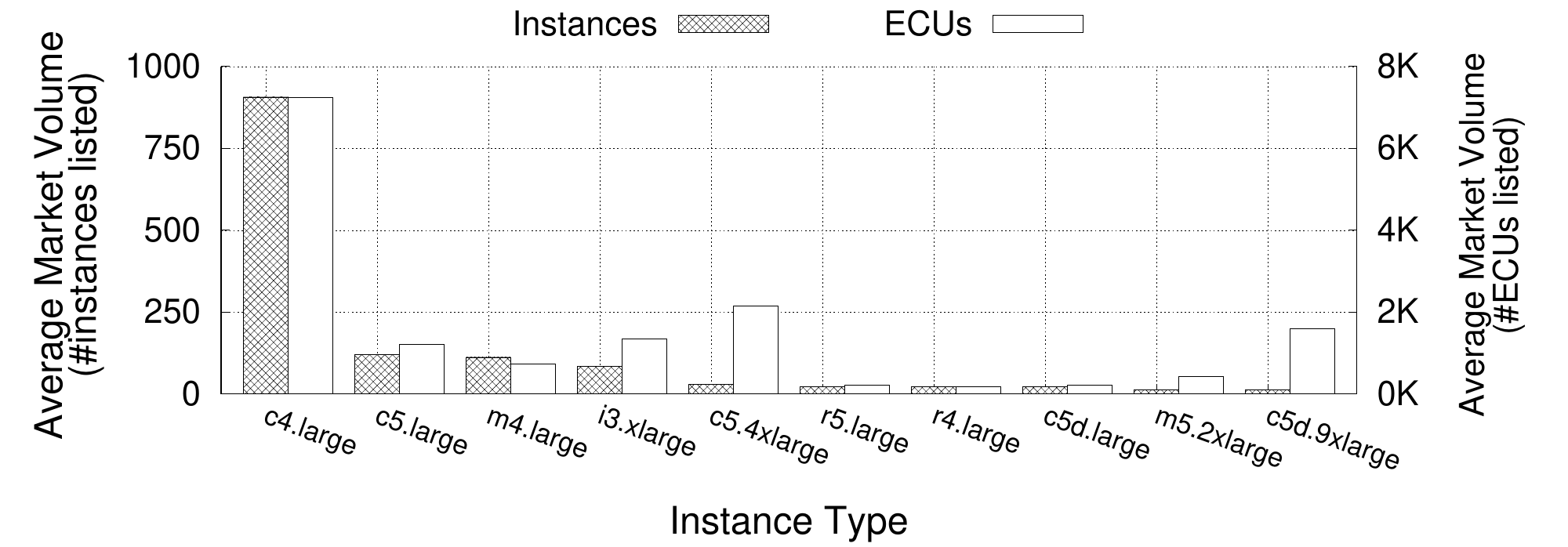}
\vspace{-0.7cm}
\caption{\emph{Average volume for different VM types in us-east-1.} }
\vspace{-0.6cm}
\label{fig:compare-types}
\end{figure}

The graph shows that there are significant differences in RIM volume both between regions and over time.  The us-east-1 had the most volume on average ranging from 16k ECUs to nearly 35k ECUs, although us-west-2's volume exceeded it during the latter part of 2019 and early 2020.  This was primarily due to one user listing 500 {\tt c5.2xlarge} RIs, which translates to $19.5$k ECUs.  This demonstrates that the market volume is still small enough that a single user can substantially affect it.  In this case, the late 2019 spike in us-west-2 increased the volume by over 3$\times$ from $\sim$10k to $\sim$30k ECUs.  There were similar spikes in us-east-1. Interestingly, us-east-2 and us-west-1 had consistently lower volume with less than 1k ECUs for the first half of 2019.  Both regions' volume was also much less volatile, with us-west-1's volume experiencing many fewer spikes than the other regions. Notably, there are no clear effects due to COVID-19 after 2020/3. 

Figure~\ref{fig:compare-regions} plots the average market volume in terms of number of VMs listed (left y-axis) and number of ECUs listed (right y-axis) for different regions.  The graph shows that us-east-1 and us-west-2 are by far the largest regions, while eu-west and eu-central are between the size of us-east-2 and us-west-1. The other regions have much lower listing volume, likely because they have less capacity.  Note the difference between the instance listing and ECU listing range on each y-axis.  While us-east-1 has over 1000 instances listed, most of the regions have only a few dozen.  This low market volume means using the RIM may be more risky in these regions with fewer RIs for sale and less diversity of types and terms. Finally, Figure~\ref{fig:monthly-sales} estimates monthly sales for the two largest regions (us-east-1 and us-west-2), as well as Amazon's 12\% cut, by assuming every listing that goes off the market is sold.   The graph shows some seasonality, particularly in us-east-1, with more sales over summer and before the winter holidays.

\noindent {\bf Market Volume by VM Type.} Since us-east-1 is the most active market, Figure~\ref{fig:compare-types} focuses in on the volume for the top ten most popular instance types (out of over 100) in us-east-1.  We can see that the {\tt c4.large} is by far the most popular type followed by the {\tt c5.large}, which represents the next generation of the same type.  The RIM enables users to sell reservations that were made in the past, and thus the  types are often from older generations.  Thus, the RIM may be a lagging indicator of instance type popularity.   The data also shows that the most popular types are smaller sized instances (large and xlarge) from the compute optimized ({\tt cX}) and general-purpose ({\tt mX}) families.   Thus, the RIM is especially well-suited for workloads that can accommodate these instance types.

\begin{figure}[t]
\centering
\includegraphics[width = 0.5\textwidth]{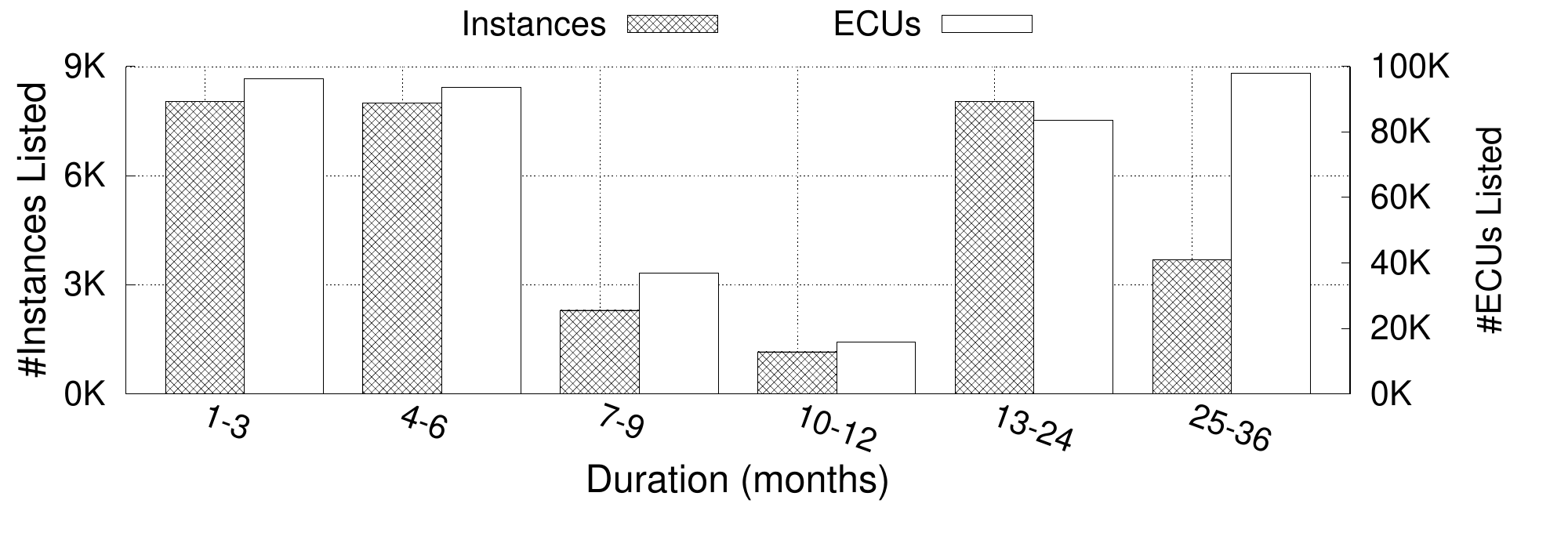}
\vspace{-1.0cm}
\caption{\emph{Listings as a function of term duration for us-east-1.}}
\vspace{-0.4cm}
\label{fig:duration-east-1}
\end{figure}

\begin{figure}[t]
\centering
\includegraphics[width = 0.5\textwidth]{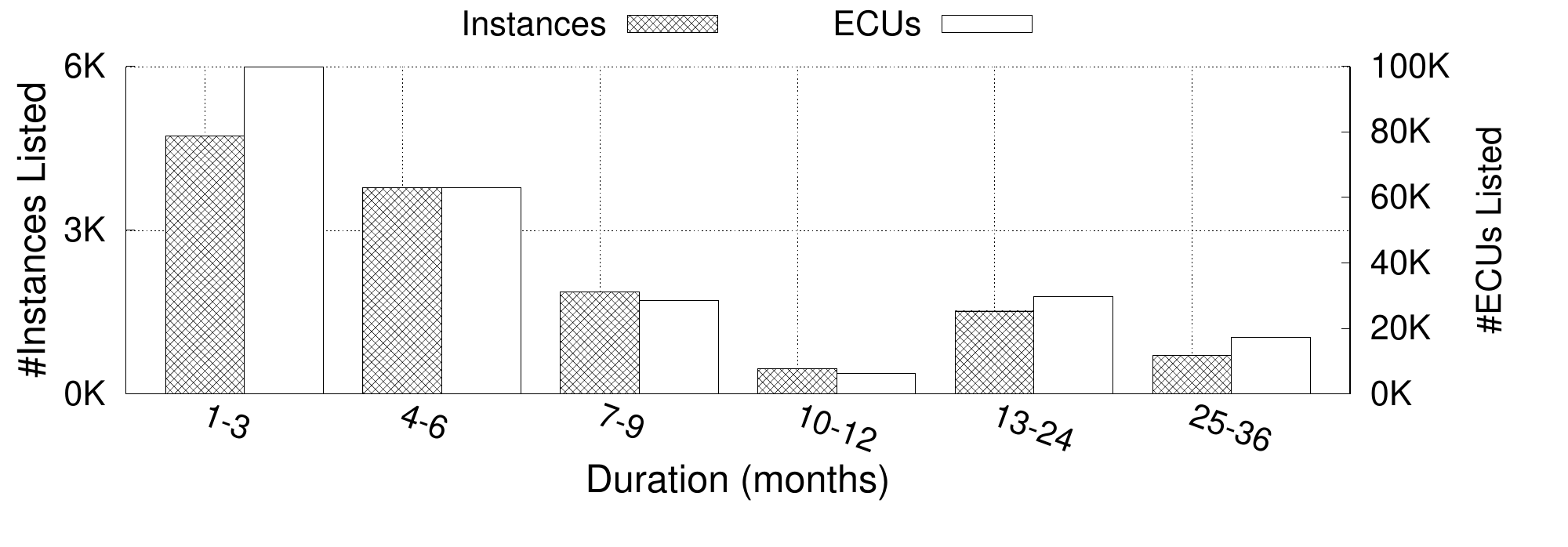}
\vspace{-1.0cm}
\caption{\emph{Listings as a function of term duration for us-west-2.}}
\vspace{-0.7cm}
\label{fig:duration-west-2}
\end{figure}

\noindent {\bf Listing Term Durations}.  Figure~\ref{fig:duration-east-1} again focuses on us-east-1 and shows the total number of listings with a specified remaining term.   Note that the duration ranges on the x-axis are not uniform.  The graph shows that terms with short durations (1-6 months) are more plentiful than listings with longer durations near 1 year (7-12 months).  This is intuitive, since users can purchase 1-year reservations from Amazon directly.  Thus, any RI with near a 1-year duration must either be a 1-year RI that a user immediately decided to sell or, less likely, a 3-year RI that has been used for 2 years.   The graph indicates that there is an opportunity to purchase these shorter-term reservations in the RIM, especially for users that can only accurately forecast their demand over months and not years.

Interestingly, there is a discrepancy in the bar height for the 25-36 month duration, which indicates a smaller number of larger sized instances on the market.  This discrepancy was the result of a large listing with 570 {\tt c4.large} instances for a duration of 19 months.   This may indicate buyer's remorse with the user of these expensive 3-year reservations trying to unload them in the market.   In addition, the large number of listings in the 13-24 month range is almost entirely due to a single listing of 1200 {\tt c4.large} instances for 14 months. Even with these large listings, though, the graph shows that there are over twice as many RIs available with durations from 1-12 months compared to these longer 13-24 month RIs. 

For comparison, Figure~\ref{fig:duration-west-2} shows the same graph for next largest region: us-west-2.  The graph shows similar trends as in us-east-1 for listings less than 1 year, with fewer listings in the 7-12 month range compared with the 1-6 month range.  However, there are significantly fewer listings with longer duration compared to us-east-1, likely due to the absence of similar massive market-moving listings as above.  Without these large listings, there is an order of magnitude fewer listings of longer durations ($>$1 year) in the us-west-2 RIM.

\begin{figure}[t]
\centering
\includegraphics[width = 0.45\textwidth]{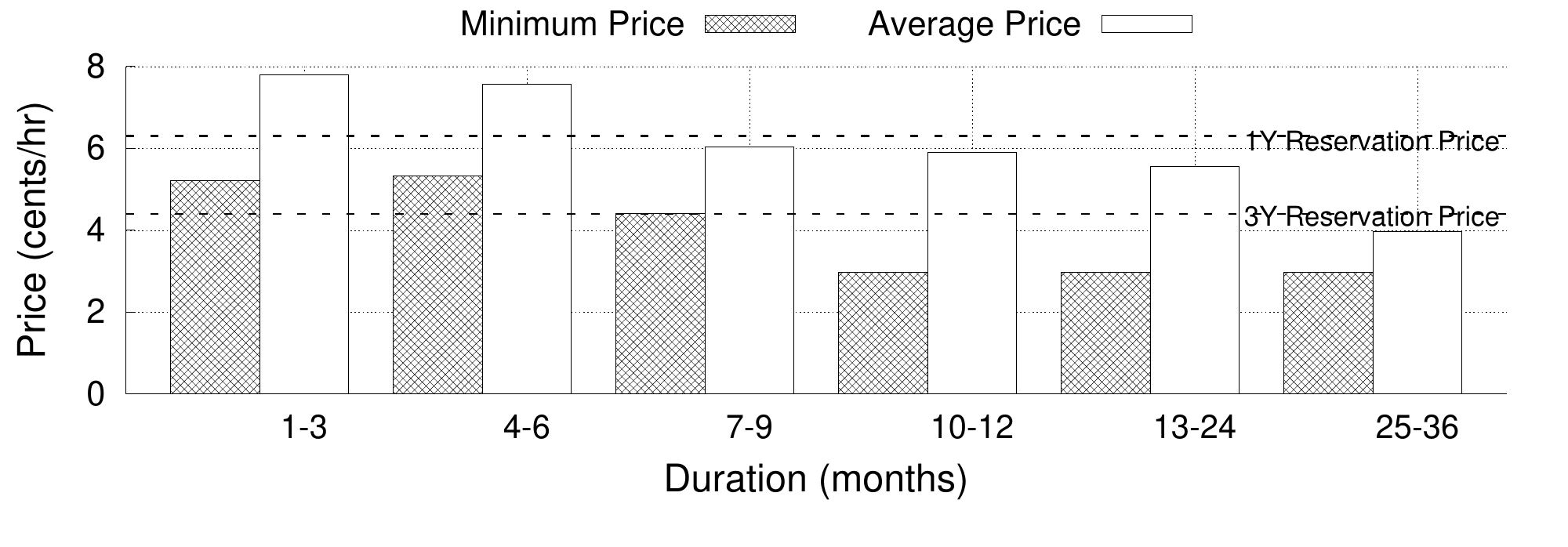}
\vspace{-0.7cm}
\caption{\emph{Price of {\tt c4.large} versus duration in us-east-1.}}  
\vspace{-0.3cm}
\label{fig:duration-c4-large}
\end{figure}

\begin{figure}[t]
\centering
\includegraphics[width = 0.45\textwidth]{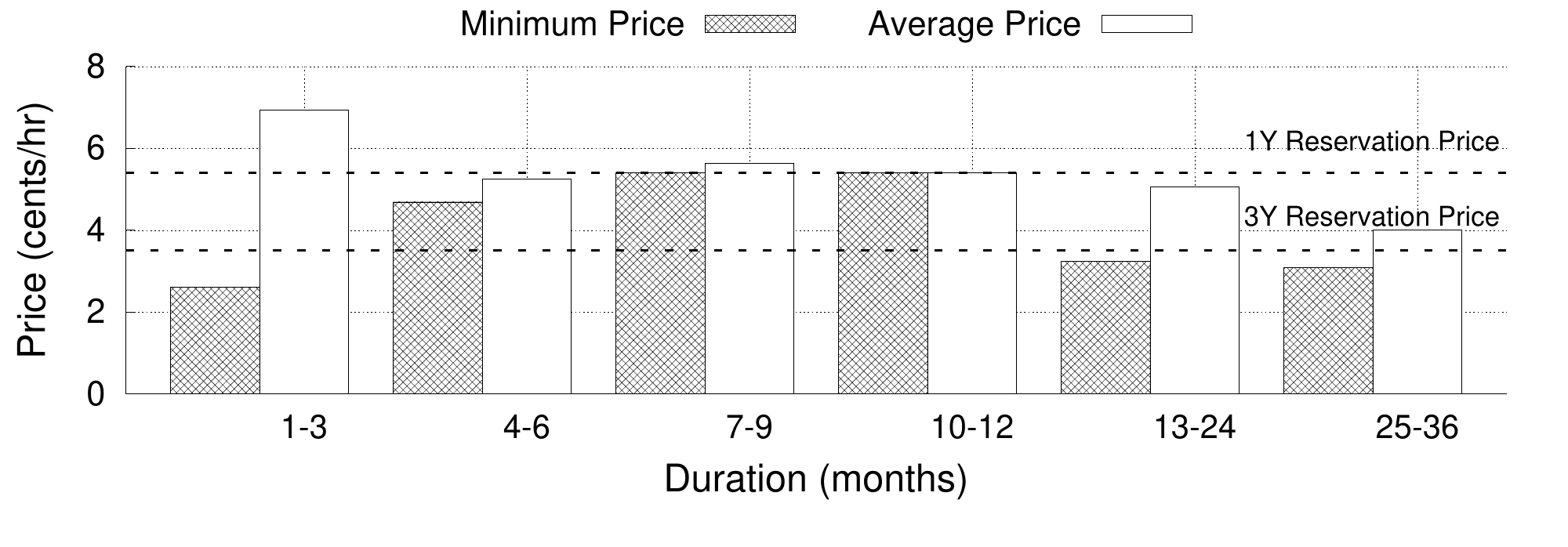}
\vspace{-0.7cm}
\caption{\emph{Price of {\tt c5.large} versus duration in us-east-1.}}  
\vspace{-0.6cm}
\label{fig:duration-c5-large}
\end{figure}

\noindent {\bf Listing Price.} Figures~\ref{fig:duration-c4-large} and \ref{fig:duration-c5-large} show the effective price of the listings for the two most popular instance types in us-east-1 ({\tt c4.large} and {\tt c5.large}).  The effective price is based on the listed upfront price amortized over the remaining term, and the recurring price.  The graph also shows dotted horizontal lines at the 1- and 3-year all upfront reserved prices. We plot the average and minimum effective price across all listings, since the RIM sells the lowest priced instances first.  We observe that as the term increases, the average listing price generally decreases in a near linear fashion.  This is intuitive, since longer reservations impose a higher risk on the buyer, which is factored into the price.  Amazon provides a higher discount for 3-year versus 1-year reservations for a similar reason.  The minimum price also follows this trend, but is slightly more volatile as it is dictated by the single user with the minimum price.  In some cases, the minimum price is less than Amazon's price.  For example, the minimum and average price for the 10-12 month term of {\tt c4.large} are less than Amazon's 1-year reserved price; the same is true for the 25-36 month term price compared to the 3-year reserved price. {\em These price inversions demonstrate the RIM's opportunity for cost savings.} 

\begin{figure}[t]
\centering
\includegraphics[width = 0.5\textwidth]{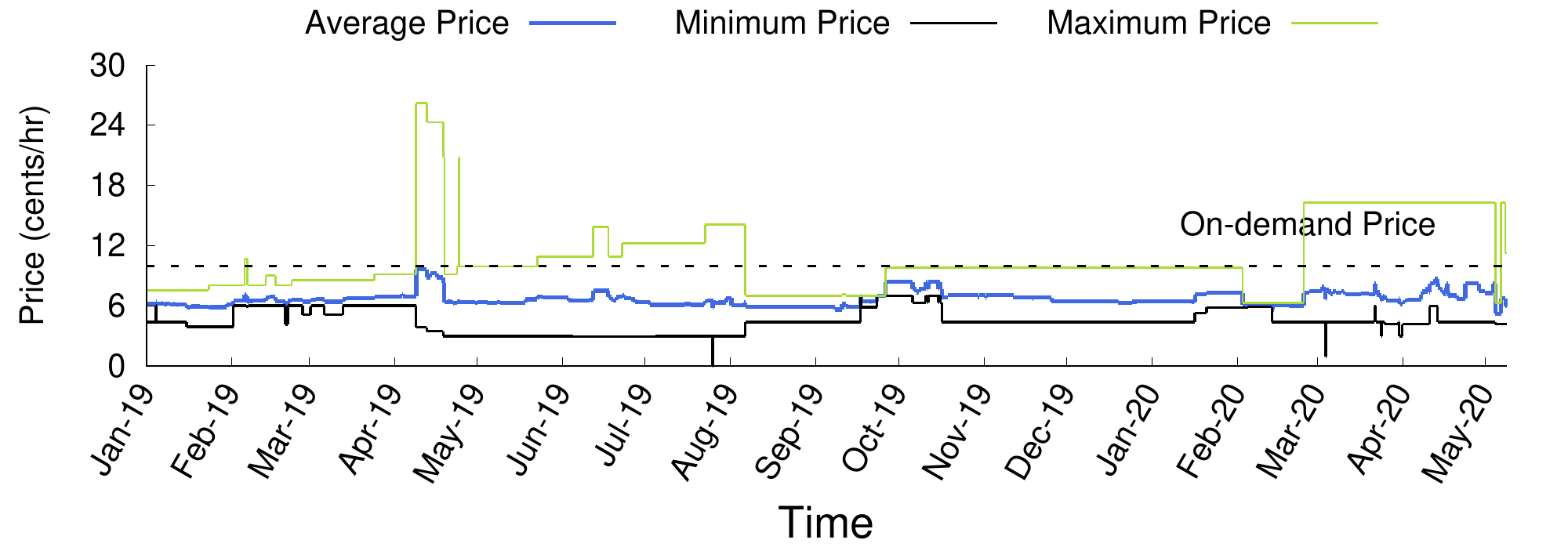}
\vspace{-0.8cm}
\caption{\emph{Price of {\tt c4.large} in us-east-1 over time.}}
\vspace{-0.3cm}
\label{fig:price-over-time}
\end{figure}

\begin{figure}[t]
\centering
\includegraphics[width = 0.5\textwidth]{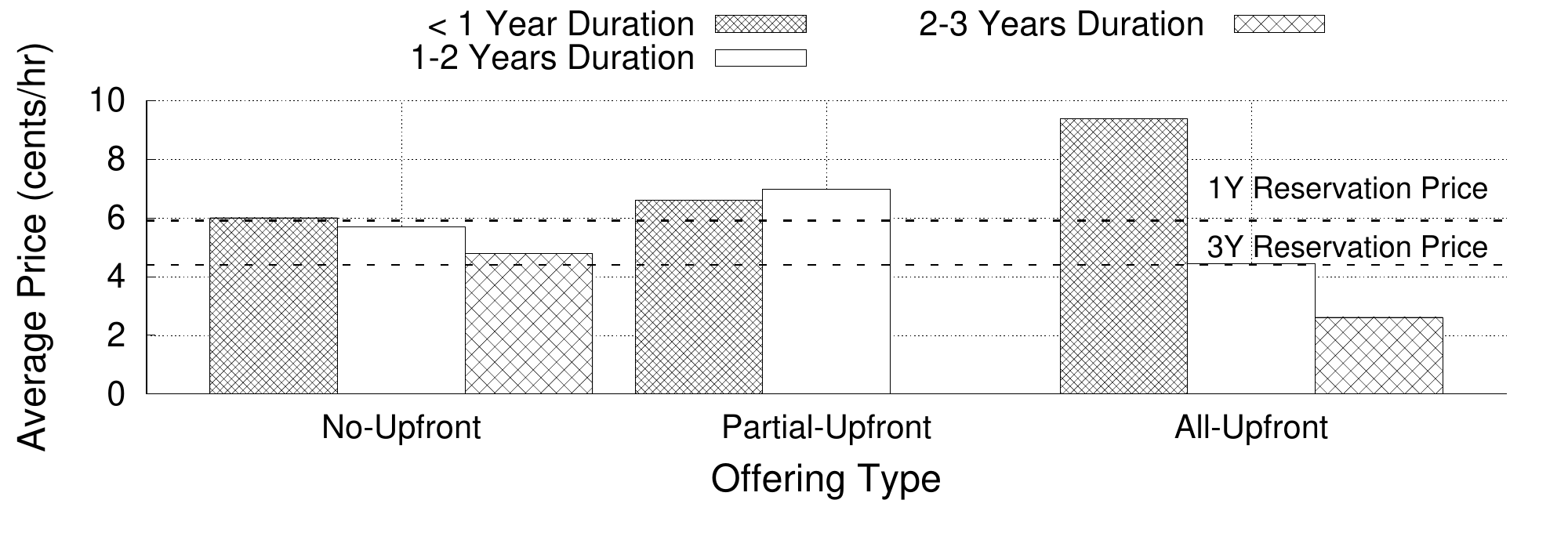}
\vspace{-1.0cm}
\caption{\emph{Prices of {\tt c4.large} listings in us-east-1 for different payment options and durations.}}
\vspace{-0.5cm}
\label{fig:compare-offering-types}
\end{figure}

\begin{figure}[t]
\centering
\includegraphics[width = 0.5\textwidth]{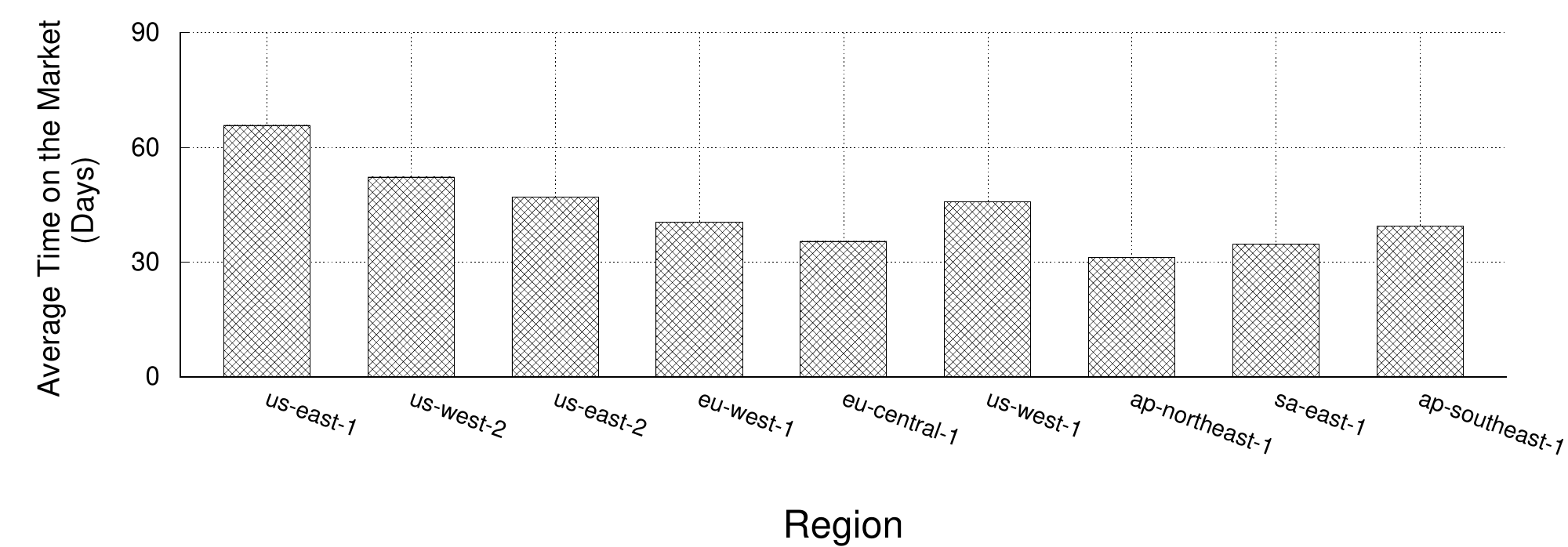}
\vspace{-0.8cm}
\caption{\emph{Time on the market for different regions.}}
\vspace{-0.4cm}
\label{fig:time-market-regions}
\end{figure}

\begin{figure}[t]
\centering
\includegraphics[width = 0.5\textwidth]{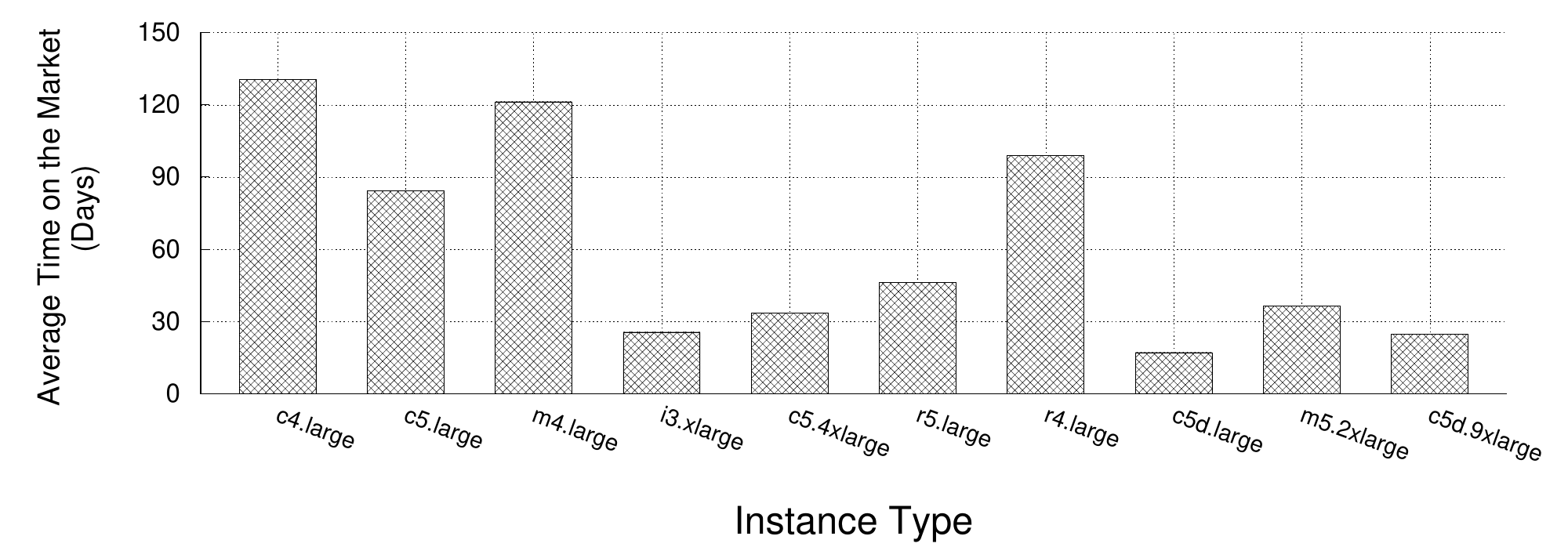}
\vspace{-0.8cm}
\caption{\emph{Time on the market for instance types in us-east-1.}}
\vspace{-0.6cm}
\label{fig:time-market-types}
\end{figure}

Figure~\ref{fig:price-over-time} shows how the listing prices, in this case for the {\tt c4.large} in us-east-1, change over time regardless of the term duration. We choose the {\tt c4.large} since it is by far the highest volume instance based on Figure~\ref{fig:compare-types}. The graph includes the minimum, maximum, and average listing price, as well as the on-demand price for the corresponding instance type.  The graph shows that the minimum and average price are relatively stable, and consistently less than the on-demand price as expected.  The maximum price periodically spikes above the on-demand price due to a few listings with exorbitantly high upfront prices. These high upfront prices suggest mispricing by sellers that deserves additional attention. 

Figure~\ref{fig:compare-offering-types} shows the average effective price of the {\tt c4.large} for different payment options (all upfront, partial upfront, and no upfront) and durations.   Recall that the effective price includes both the upfront price amortized over the listing duration and the recurring charges.  We would expect the no upfront price to be the highest for any given duration range, since its original selling price is the highest.  Interestingly, the no upfront option has the lowest price for short durations ($<$1 year), and has the second lowest price for medium durations (1-2 years).  Since nearly all no upfront RIs are \$0, and thus represent only a short-term capacity reservation, it may be that users are pricing them relative to on-demand capacity reservations, which are 40\% more expensive. 

The no upfront option with short duration is nearly equal to the amortized 1-year reserved price, while the partial upfront and all upfront options are much more expensive.  This may again represent mispricing by sellers that deserves additional attention. In particular, the partial upfront option is likely the most complex to price (given it has both an upfront and recurring charge) and appears to be the most mispriced, with both the short and medium term durations having higher average effective prices than the 1- and 3-year reservations. In contrast, the long term (2-3 year) all upfront option is most discounted relative to the 3-year amortized reserved price at nearly half the price, {\em thereby offering cost saving opportunities.}

\noindent {\bf Time on the Market}.  Figures~\ref{fig:time-market-regions} and \ref{fig:time-market-types} show the average time on the market for multiple regions and the highest volume instance types in the us-east-1, respectively.  We order bars by volume as in Figures~\ref{fig:compare-regions} and \ref{fig:compare-types}, respectively.  However, while volume is an indication of supply, time on the market is an indication of demand relative to that supply.  


%
As Figure~\ref{fig:time-market-regions} shows, the regions with the largest volume tend to also have the longest average time on the market with an average of over 2 months for us-east-1.  This indicates that demand is less relative to the supply in the larger RIM markets.  Space constraints preclude a more in-depth analysis, but clearly a follow-up analysis would show whether the law of supply and demand holds across regions, i.e., that prices rise as demand increases relative to supply and vice versa.  Figure~\ref{fig:time-market-types} also shows that the instance types with the largest supply do not have the lowest time on the market, i.e., highest demand.  For example, the largest volume {\tt c4.large} type has the longest time on the market---over 4 months---with the next highest volume types have a similarly long time. With a couple of exceptions the instance types with lower volumes spend much less time on the market, often less than 30 days. 

\begin{figure}[H]
\centering
\includegraphics[width = 0.5\textwidth]{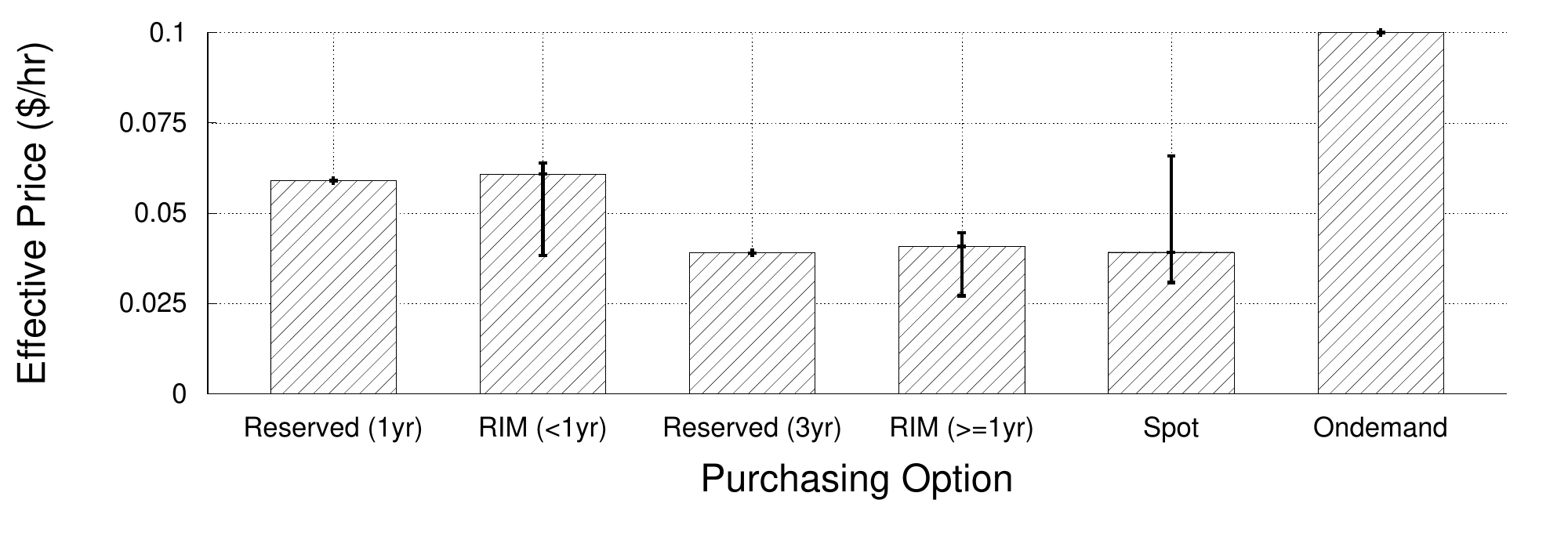}
\vspace{-1cm}
\caption{\emph{Effective prices for the c4.large in us-east-1.}}
\vspace{-0.4cm}
\label{fig:spot-cmp}
\end{figure}

\noindent {\bf Comparing Purchasing Options.}  Figure~\ref{fig:spot-cmp} shows the effective price of different purchasing options from 02/2020 to 05/2020 for the {\tt c4.large}. We only use 3 months of data, since spot prices are accessible for only the past 3 months.  For the RIM, we plot the effective price of RIs sold based on their term duration, selling price, and recurring charges.  For the RIM and spot prices, we plot error bars representing the 5th and 95th percentile price.  As expected, the graph shows that on-demand VMs have the highest price.  The graph also shows that users neither lose nor make money on average in the RIM, as its price is similar to the 1- and 3-year reserved option.  However, the RIM does exhibit variance, with some users receiving large discounts and some incurring losses. The RIM for 3-year reservations also has a similar average price as spot VMs, and thus they may be preferable for workloads greater than 1 month, i.e., the minimum holding period.

\vspace{-0.3cm}  
\section{Conclusion}
\label{sec:conclusion}
This paper provides a first look at data from Amazon's Reserved Instance Marketplace.  Our initial analysis motivates future work that provides a more in-depth analysis and uses it to develop intelligent strategies for exploiting the RIM to mitigate demand risk and optimize long-term cloud costs.  

\noindent {\bf Acknowledgements.} We thank the anonymous reviewers and our shepherd, Matei Zaharia, for their comments. This work is funded by NSF grants CNS-1802523 and CNS-1908536.



\section{Discussion Topics}
\label{sec:discussion}

Our paper shines a light on an interesting part of Amazon EC2 that has thus far been largely overlooked by the cloud computing research community.   While many papers have focused on EC2's spot market, there have been only a few that have focused on the RIM.  Our paper shows that the RIM is large with tens of thousands of listings offered for tens of millions of dollars. We also show that the RIM is a more complex and richer market than the spot market.  We believe the lack of research is likely due, in part, to the lack of available RIM data, which our paper attempts to address.  

This paper's data analysis represents only a first look at the RIM and its complexity, and likely raises more questions than it answers.    For example, what factors affect the time on the market and what accounts for seller mispricing?  Beyond the direct implications to using the RIM, our data also provides an indirect glimpse into Amazon's cloud, including the relative size and scale of its AZs and regions as well as its revenue (assuming each listing that disappears is sold and Amazon takes a 12\% cut). We would appreciate feedback from the workshop on the potential implications of our analysis, and additional data analyses that would be useful and informative. 

We would also appreciate hearing any audience experiences or anecdotes with using the RIM, since there is little published work on it.  Does industry commonly use RIs and the RIM to balance demand risk and cost?  Or are cloud costs small enough that such optimizations are not important?  How obscure is use of the RIM?  Ultimately, our goal is to develop data-driven strategies for intelligently using the RIM to cost-effectively provision large-scale cloud infrastructure.  Thus, we would like to hear about how people are currently using the RIM, and thoughts on how our data analysis might affect the current strategies, if any.  Clearly, using the RIM should enable increased use of RIs, since users are no longer committed to using them over the entire term of the reservation.

\bibliographystyle{plain}
\bibliography{paper}

\end{document}